\documentclass{agujournal}

\journalname{Geophysical Research Letters}

\begin{document}

\title{Charge proportional and weakly mass-dependent acceleration of different ion species in the Earth's magnetotail}

\authors{F. Catapano\affil{1,2}, G. Zimbardo\affil{1}, S. Perri\affil{1}, A. Greco\affil{1}, D. Delcourt\affil{2}, A. Retin\`{o}\affil{2}, I. J. Cohen\affil{3}}

 \affiliation{1}{Dipartimento di Fisica, Universit\'{a} della Calabria,Rende (CS), Italy}
 \affiliation{2}{Laboratoire de Physique des Plasmas, Universit\'{e} Pierre et Marie Curie, Paris, France}
\affiliation{3}{Johns Hopkins University Applied Physics Laboratory, Laurel, Maryland, USA}

\correspondingauthor{Filomena Catapano}{filomena.catapano@unical.it}

\begin{keypoints}
\item Test particle simulations of the acceleration of different ion species by stochastic fluctuations in the Earth's magnetotail are performed.
\item For the top 10\% energetic ions, the energy gain grows almost linearly with the ion charge, while the dependence on the ion mass is weaker.
\item We propose that the possible presence of O$^{5+/6+}$ ions may contribute to explain the strong oxygen energization observed in the magnetotail.
\end{keypoints}

\begin{abstract}
	Energetic particles with energies from tens of keV to a few hundreds keV are frequently observed in the Earth's magnetotail. Here we study, by means of a test particle numerical simulation, the acceleration of different ion species (H$^{+}$, He$^{+}$, He$^{++}$, and O$^{n+}$ with $n=1$--$6$) in the presence of transient electromagnetic perturbations. All the considered ions develop power-law tails at high energies, except for O$^+$ ions. This is strongly correlated to the time that the particle spend in the current sheet. Ion acceleration is found to be proportional to the charge state, while it grows in a weaker way with the ion mass. We find that O$^{5+/6+}$ can reach energies higher than $500$ kev. These results may explain the strong oxygen acceleration observed in the magnetotail.

\end{abstract}

\section{Introduction}

Understanding the acceleration mechanisms of energetic ions and electrons in the geospace environment is one of the main outstanding issues of space physics. In the Earth's magnetotail, energetic particles with energies from tens of keV to a few hundreds keV are found \citep{Christon89,Keiling04,Imada07,HaalandEA2010,Artemyev14}; in particular, heavy ions, like oxygen ions, are observed in the magnetotail with energies often reaching several hundreds keV \citep{Keika13,Luo14,Kronberg14, Kronberg15}. Spacecraft observations suggest that both electrons and ions are accelerated not only in the vicinity of the reconnection $X$-line, but also in a larger area around the reconnection region \citep[e.g.,][]{Imada07,Ono09}, possibly by a process related to time dependent patchy reconnection \citep{Grigorenko09}.
An acceleration mechanism, that has been widely investigated, is that occurring at dipolarization fronts due to reconnection jets in the geomagnetic tail  \citep [e.g.,][]{Ashour-Abdalla11, Ukhorskiy13, Birn14, Greco14, Greco15, Greco17}. 

Observations also suggest that heavy ions are accelerated differently from protons, depending on their mass and also on the acceleration mechanism.
Typical analytical estimates of the ion acceleration in the current sheet due to the dawn-dusk electric field yield a mass proportional energy gain \citep[e.g.,][]{Zelenyi07,Kronberg14}. Furthermore, the acceleration of ions in plasmoids associated with magnetic and electric field fluctuations was found to be more effective for heavy ions \citep{Grigorenko15}.
 However, recent Geotail data analysis by \citet{Ohtani15} showed that for fast earthward flows, corresponding to local dipolarizations, the energy density increase is not mass dependent. \citet{Ono09} investigated the variation of the energy spectra of protons and oxygen ions during dipolarization events. It was found that the ions are non-adiabatically accelerated by the magnetic fluctuations and that this mechanism is sometimes more efficient for protons than for oxygen ions.

In order to shed light on this complex variety of observations, we carry out test particle simulations in which protons and other ion species are injected in the magnetic field configurations obtained by \citet{Catapano15}, corresponding to a class of solutions that generalize the Harris model. The three dimensional (3D) stochastic electromagnetic perturbations of \citet{Perri11} are included in the model; these perturbations may correspond to the random transient flows and electric and magnetic fluctuations which are frequently observed in the magnetotail \citep[e.g.,][and references therein] {Cattel82,Borovsky97,Borovsky03,Zimbardo10}. The role of the magnetic fluctuations in the acceleration process has been pointed out by \citet{Ono09,Nose14}. Recently, \citet{Catapano16} studied the acceleration of H$^+$, He$^+$ and O$^+$ while varying the equilibrium magnetic field profile and the intensity of the perturbations. Also, high charge state ions are observed in the magnetosphere \citep{Kremser87}. Recent analysis of the Polar spacecraft data has shown that a substantial population of O$^{6+}$ enters the magnetosphere from the solar wind \citep{Allen16a,Allen16b}. The O$^{6+}$ charge exchanges into O$^{5+}$, O$^{4+}$, and O$^{3+}$ while drifting towards low $L$ shells. The radial and local time distribution of these ions are shown in Figures 2 and 3 of \citet{Allen16a}, with O$^{5+/6+}$ being second in abundance to O$^{+}$ in the radial range from 10 to 20 $R_E$, and with O$^{++}$ of probable ionospheric origin coming third in abundance. Therefore, it is important to study the acceleration of such high charge state oxygen ions.

\section{Numerical Model}

Test particle simulations are performed using an electromagnetic field model for the magnetotail, composed by two parts: an unperturbed equilibrium configuration and three-dimensional (3D) stochastic electromagnetic fluctuations \citep{Perri11, Catapano16}. While Harris magnetic field profile is widely used in space plasmas, many observations in the Earth's magnetotail show that the current sheet (CS) has a non-Harris-like profile, with the CS being often embedded in a thicker plasma sheet \citep[e.g.,][]{Runov06,Petru15}, and this property has also been used for numerical simulations \citep[e.g.,][]{Malova13}. Recently, \citet{Catapano15} have generalized the Harris current sheet to the case when several current carrying populations, i.e., multiple electron and ion populations, are present. This model allows to adjust the level of plasma temperature and density inhomogeneities across the CS in a wide range of configurations, with the magnetic field profile being obtained self-consistently (see \citet{Catapano15} for more details).
\citet{Catapano16} studied the ion acceleration while changing the parameters of the model like the equilibrium magnetic field profile and the strength of the perturbations. However, \citet{Catapano16} have found that the  details of the magnetic field profile $B_x(z)$ have little influence on the energization process, so that for further investigations we will use only one solution for the magnetic field profile, in particular the solution for which the ratio between electron and proton temperature is $1 / 3$ .

We consider a coordinate system in which the $x$ axis is directed along the Sun-Earth direction, the $z$ axis is oriented perpendicular to the CS and the $y$ axis is directed along the electric current flow. We investigate the Earth's middle tail, approximately from 15 to 30 $R_E$, using a three-dimensional simulation box with side lengths given by $L_x = L_y = L = 10^5$ km and $L_z= 2.5 \times 10^4$ km; the box size along the $z$ direction ranges from $-L_z$ to $L_z$. The box size in the $(x,y)$ plane is of about 15 $R_E$ which is only a fraction of the actual magnetotail extension, so that large-scale variations of the equilibrium configurations are neglected. The characteristic half-width of the CS is set to $\lambda = L_z/5= 5\times 10^3$ km $\sim 0.8$ $R_E$ \citep{Catapano15, Catapano16}. 
We superimpose on the self-consistent equilibrium solution the time-dependent electromagnetic fluctuations, as modeled in \citet{Perri11}. These represent isolated electromagnetic structures frequently observed in the magnetotail \citep{Borovsky97, Borovsky03}. The fluctuations are introduced by the perturbed vector potential 
\begin{linenomath*}
\begin{eqnarray}
A_x(\textbf{r}, t) &=& A_0 \Sigma_i \exp [- |\textbf{r}-\textbf{r}_i(t)|/ \ell] \nonumber \\
A_y(\textbf{r}, t) &=& A_0 \Sigma_i (-1)^i \exp [- |\textbf{r}-\textbf{r}_i(t)|/ \ell]
\label{eq1}
\end{eqnarray}
\end{linenomath*}
where $\textbf{r}$ is the position of a particle, $\textbf{r}_i(t)$ is the position of the fluctuation center at time $t$ and the sum is made on the number of perturbations $i$. The parameter $\ell$ represents the spatial scale of the vector potential and $A_0 / \ell$ is the peak amplitude of perturbed magnetic field and it is typically $10$ nT, as it is often observed \citep{Borovsky97}. The positions $\textbf{r}_i$ are fixed randomly in the $(x,y)$ plane and are oscillating with velocity $V=400$ km s$^{-1}$, that is a typical value for the Alfv\'{e}n speed in the magnetotail \citep{Hoshino94, Nakamura04, Voros07}, with random phases and with a typical frequency that is $\omega = V / \ell$. The amplitude of the oscillations also equals $\ell$. It can be noted that the term $(-1)^i$ in the second line in Eq. \ref{eq1} leads to $y$-components of fluctuations with alternating sign. The average separation scale between the fluctuations is $2\ell$, so that when a smaller value of $\ell$ is used, a larger number of isolated structures is present in the simulation box. The equations for the electric and magnetic fields are 
\begin{linenomath*}
	\begin{eqnarray}
	\textbf{E}(\textbf{r}, t) &=& E_{0y} \textbf{e}_y + \delta \textbf{E} (\textbf{r}, t) \nonumber \\
\textbf{B}(\textbf{r}, t) &=&  B_x (z) \textbf{e}_x + B_{n} \textbf{e}_z +\delta \textbf{B} (\textbf{r}, t) .
	\label{eq2}
	\end{eqnarray}
\end{linenomath*}
The perturbed terms in Eq. (\ref{eq2}) for the electric and magnetic fields are obtained as $\delta \textbf{E}(\textbf{r},t)=\partial\textbf{A}(\textbf{r},t)/\partial t$ and $\delta \textbf{B}(\textbf{r},t) = \triangledown \times \textbf{A}(\textbf{r},t)$, where $\textbf{A}(\textbf{r},t)$ is the vector potential in Eq. (\ref{eq1}). The perturbed electric field $\delta \textbf{E}$ has components only in the $(x,y)$ plane. We further notice that $\delta B_z$ has even parity in $z$, while $\delta B_x$ and $\delta B_y$ have odd parity in $z$ and go to zero for $z=0$; therefore, the magnetic perturbations represent moving magnetic islands. 
In addition, the term $B_x(z)$ corresponds to the magnetic field coming from the equilibrium solutions of Eq.~(3) of \citet{Catapano16}. Other terms are the constant dawn-dusk electric field component $E_{0y}=0.2$ mV m$^{-1}$, and the constant out-of-plane magnetic field component $B_n=3$ nT \citep{Sergeev03}. All the equations are normalized using the following quantities : $L=10^5$ km, $B_0 = 2$ nT, $t_0 = \Omega_{0p}^{-1} = 5$ s, $E_0= B_0 L / t_0 = 40$ mV m$^{-1}$ and the asymptotic value of the magnetic field is $B_{0x}= 20$ nT.
 An ensemble of $10^4$ particles is simultaneously injected in the simulation box at $t=0$, at $z=0$ and randomly distributed in the $(x,y)$ plane. The initial energies $E_i$ for protons and heavy ions are extracted from a Maxwellian distribution with thermal energy $E_{inj}$; heavier ions are injected with smaller velocities in order to have the same initial energies of protons. Particles exiting the simulation box are replaced by freshly injected ones, which are extracted from the initial energy distribution in order to get a long-time, steady-state solution. We numerically solve the equations of motion using as integration step $\Delta t = 0.001$ $\Omega_{0p}^{-1}$.

 \section{Numerical Results}

\begin{figure}
 	\centering
 	
 	\includegraphics[width=30pc]{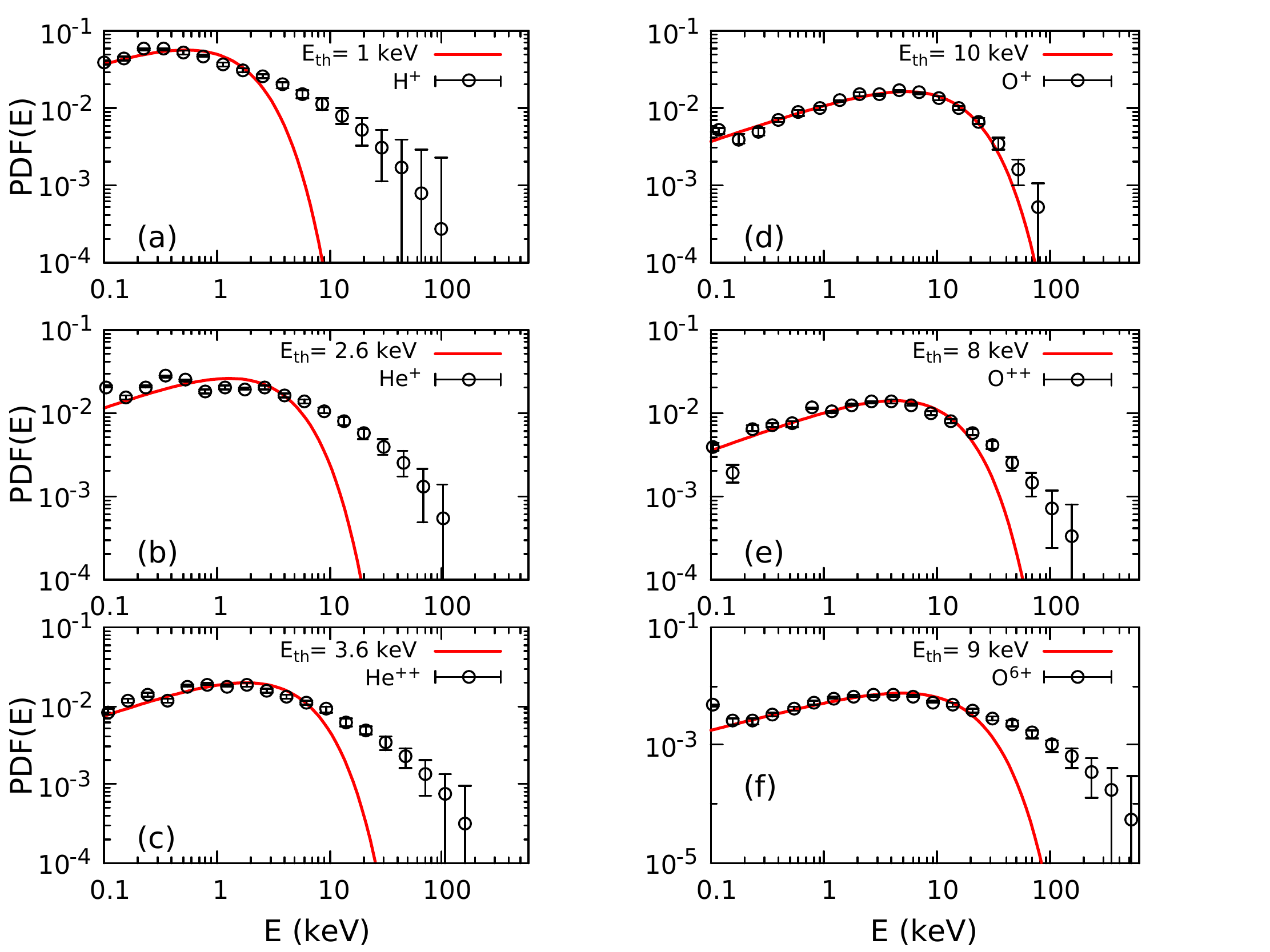}
 	\caption{Panels show the final PDFs as a function of energy (black circles) for (a) H$^+$, (b) He$^+$, (c) He$^{++}$, (d) O$^{+}$, (e) O$^{++}$ and (f) O$^{6+}$ fitted by a Maxwellian distribution (red solid lines) in the case with $\ell=8000$ km and $E_{inj} = 1$ keV.}
 	\label{fig1}
 \end{figure}

We explore the acceleration of proton and heavy ions like He$^+$, He$^{++}$ and O$^{n+}$ with $n=1$--$6$. We set the initial energy equal to $E_{inj} = 1$ keV for all species and the size of fluctuations $\ell = 8000$ km, that is the typical observed scale of the bursty bulk flow \citep[e.g.,][]{Borovsky03,Nakamura04}. In this case the oscillation frequency of the perturbations is $\omega= 0.05$ $s^{-1}$, which is smaller than the proton gyrofrequency in the CS, that is $\Omega_p= 0.3$ $s^{-1}$, and larger than the O$^+$ gyrofrequency of $\Omega_{O+}= 0.0187$ $s^{-1}$. Figure \ref{fig1} depicts the Probability Density Function of the particles energy (PDF(E)) for protons and heavier ions (black circles), fitted by a Maxwellian distribution (solid red line) characterized by a thermal energy $E_{th}$. We can notice that singly charged ions (H$^+$, He$^{+}$ and O$^+$) reach maximum energies up to 100 keV, whereas ions with higher charge states reach significantly larger energies (see for example the case of O$^{6+}$ ions in panel (f)). This shows that the random electromagnetic perturbations form an efficient acceleration mechanism for ions in the magnetotail, as suggested by observations \citep{Nose14}. All the results presented in this section have also been obtained by varying the initial energy $E_{inj}$, the size of the fluctuations $\ell$ and so their frequency $\omega$, looking for resonance effects (not shown). However, we found that there is no evidence of resonant acceleration, that is, for all ion species the energy gain increases with the size of the perturbation (which is inversely proportional to the perturbation frequency) in a rather smooth, species-independent way.
	We can conclude that the parameters $E_{inj}$, $\ell$ and $\omega$ do not affect the acceleration process in a significant way.

 It can be seen that the O$^+$ distribution function is very close to Maxwellian (panel (d)), whereas the other ions develop non thermal power-law tails. To gain insight into the origin of the different PDF shapes, we evaluate the fraction of time that ions spend in the CS. Indeed, ions may interact with the fluctuations as in a stochastic Fermi process \citep{Perri09, Greco10}, so that the shape of the final PDF(E) is related to the time that the particles spend in the acceleration region, i.e. within the CS (defined by $|z| < \lambda$) where the fluctuations are most intense and particles less magnetized. 

\begin{figure}
	\centering
	
	\includegraphics[width=25pc]{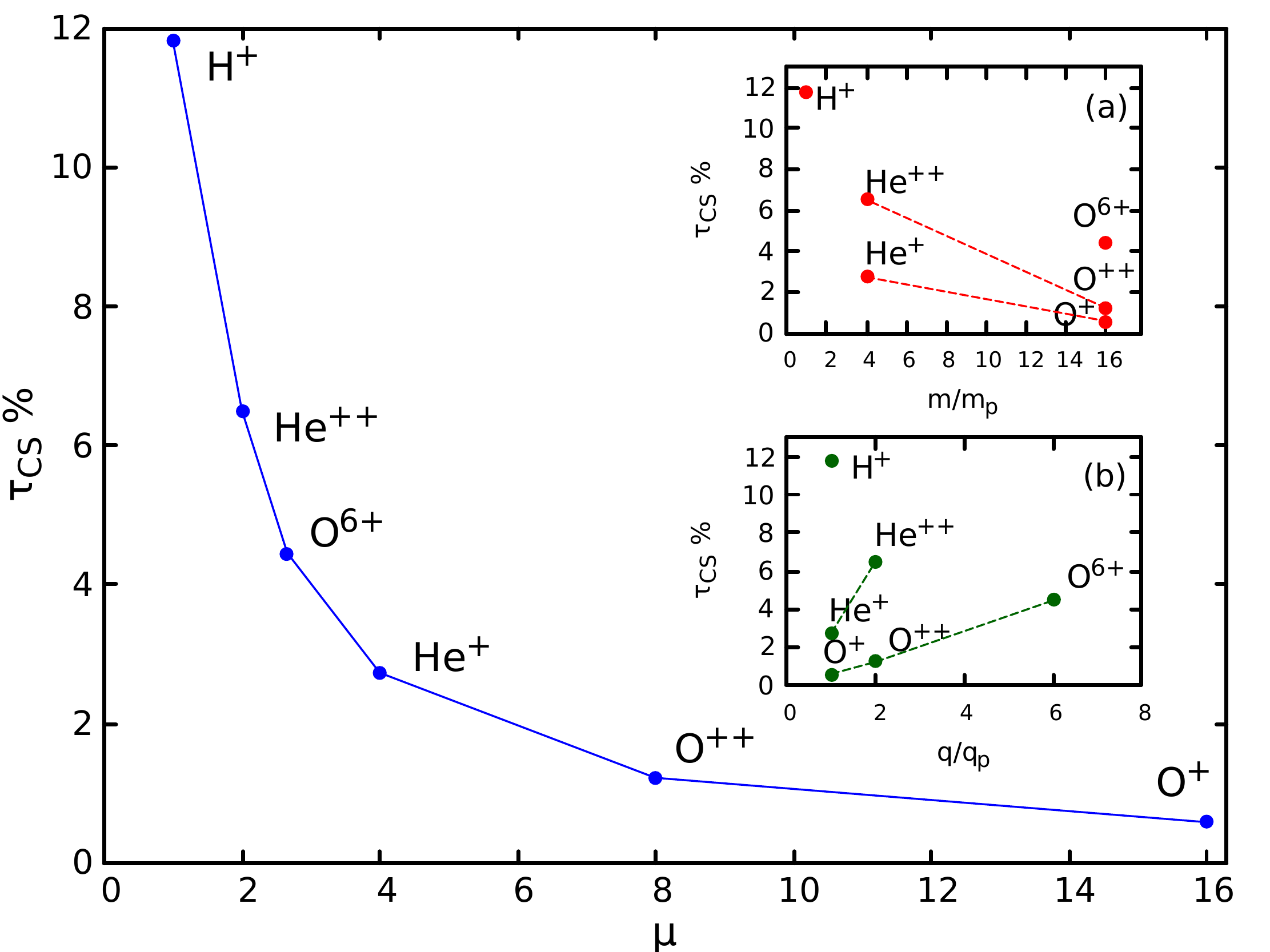}
	\caption{Residence time in the CS for different ion species as a function of the normalized mass-to-charge ratio $\mu$. The insets show the residence time as a function of (a) mass and (b) charge state normalized to those of protons. The residence time is given in percentage with respect to the total residence time in the simulation box. The red lines in panel (a) connect the ions with equal charge state, while green lines in panel (b) connect ions with equal mass.}
	\label{fig2}
\end{figure}

Figure \ref{fig2} shows the percentage of time $\tau_{CS}$ that the ions spend in the CS as a function of the normalized mass over charge ratio $\mu = (m/m_p)/(q/q_p)$. The parameter $\tau_{CS}$ is evaluated with  respect to the total residence time of the ions in the simulation box. We can see that $\tau_{CS}$ is well  ordered with $\mu$. Clearly, this is due to the fact that $\mu$ is proportional to the Larmor radius (see also trajectories in Figure 15 of \citet{Kronberg14} and in Figure 5 of \citet{Catapano16}). Therefore, ions that spend a long time in the CS interact many times with the random fluctuations and undergo a Fermi-like process, thus developing the power-law tail. Conversely, O$^+$ ions spend a limited time in the CS, so that the resulting PDF(E) remains nearly Maxwellian. The power-law tails have been fitted and the best-fits give a slope that decreases (flatter tails) as $\tau_{CS}$ increases, in agreement with stochastic Fermi acceleration \citep{Fermi49}.

 From the insets in Figure \ref{fig2} it is possible to note that $\tau_{CS}$ has not a unique trend with the ion mass or charge  separately (panels (a) and (b)). Following the red lines in panel (a), we can see that the residence time decreases with the ion mass. In a similar way, following the green lines in panels (b), we find that the residence time increases with the charge state. These results show that the residence time depends on the ion mass for a specific charge and on the charge for a specific mass. It is also worth noting that the maximum energy of the ions reported in Figure \ref{fig1} does not increase with the residence time in the current sheet, as found by \citet{Artemyev09}, too.

\begin{figure}
	\centering
	
	\includegraphics[width=30pc]{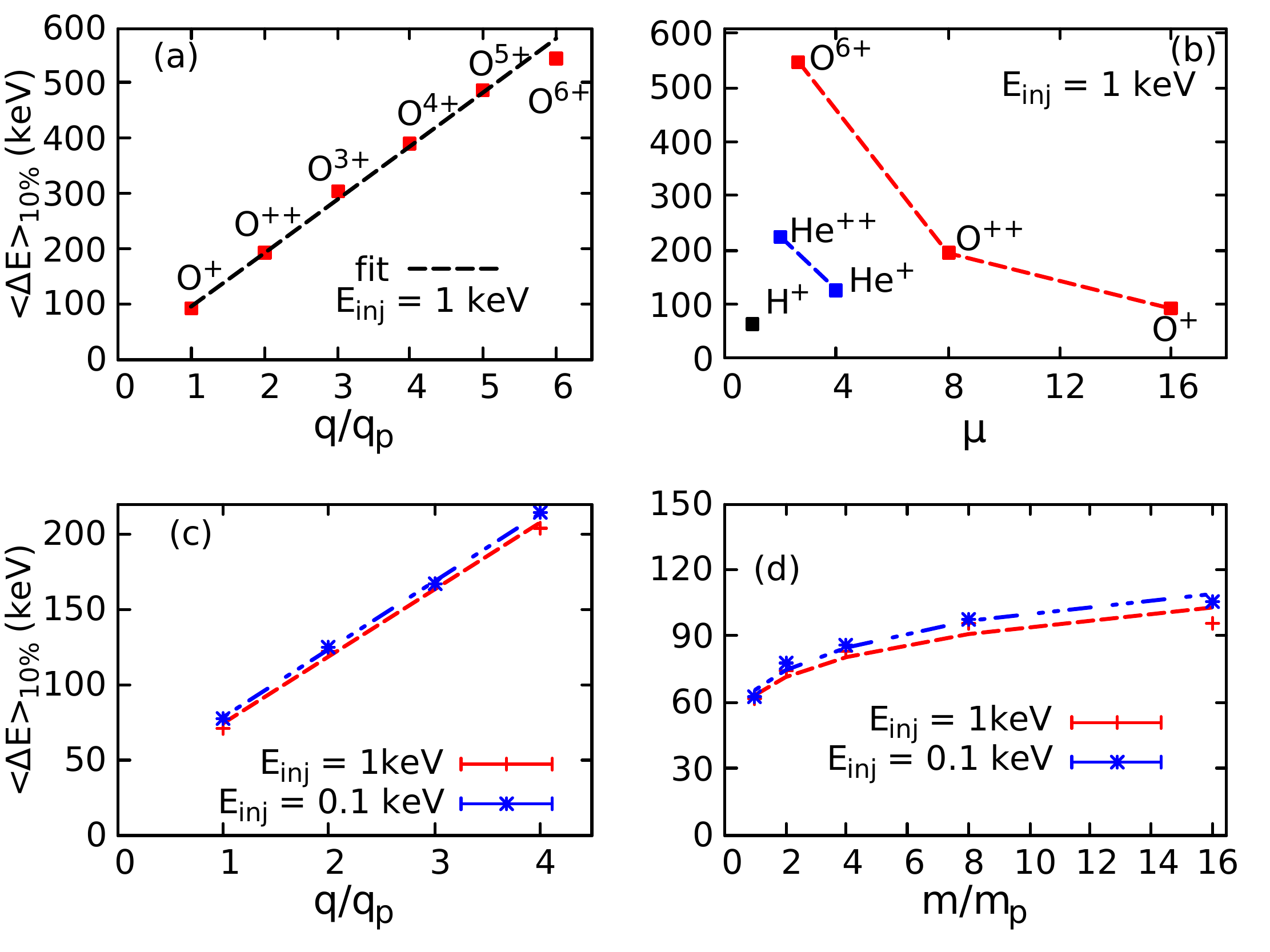}
	
	\caption{Energy gain of the top 10\% energetic particles as function of (a) charge and (b) mass-to-charge ratio for the case with $E_{inj} = 1$ keV. In panels (c) and (d) the same quantity is given for ad-hoc particles (see text for details) as a function of their charge state and mass, respectively. The red (blue) points represent the case with $E_{inj} = 1$ keV ($0.1$ keV). Dashed lines are used to show the best fit, while the thin dashed lines in the panel (b) connect ions with equal mass.}
	\label{fig3}
\end{figure}

To quantify the acceleration dependence on the charge and mass, we consider the 10\% of the test particle population that experience the largest energization $\Delta E = E_f - E_{i}$, where $E_f$ is the final energy and $E_{i}$ is the initial energy of each particle. Figure \ref{fig3} shows the average energy gain $\langle\Delta E\rangle_{10\%}$  as a function of the charge (panels (a) and (c)), mass over charge ratio (panel (b)) and mass (panel (d)) for different ions. First we consider the case with $E_{inj} = 1$ keV and ions O$^{n+}$ (Figure \ref{fig3}(a)). We can see that the energy gain grows with the ion charge. The best fit (black dashed line) shows that the energization linearly depends on the charge state, and the slope was found to be 92 keV/q with a value of $ \chi^{2} \sim 0.008$.  Panel (b) displays the same quantity for different ions as a function of the parameter $\mu$. It can be noted that the energy gain is not ordered in a simple way as a function of $\mu$ for all ion species. However, from this panel it is clear, once again, how the energization increases with the charge state at a fixed mass (dashed lines). In order to further investigate the role of ion mass and charge, we performed further runs introducing virtual particles. Testing the model with ad-hoc particles is a tool to better investigate the dependence on ion mass and charge. In Figure \ref{fig3} (c) we report the simulation  results for ions with constant mass $m_i=m_p$ and charge ratio $q_i /q_p = $ 1; 2; 3; 4; while in Figure \ref{fig3} (d) for the case with constant charge $q_i =  q_p$ and different mass ratio $m_i / m_p =$ 1; 2; 4; 8; 16. We perform the simulations using injection energy equal to $E_{inj} = 1$ keV and $E_{inj} = 0.1$ keV (red and blue symbols, respectively). The lower energy corresponds to the cold ions that are sometimes observed in the magnetotail \citep{Seki03}. The linear fits (dashed lines) in panel (c) return a value of $ \chi^{2} < 0.02$ and the slope is $\sim $ 44 keV/q for both cases. We stress that this slope is smaller than the one obtained for oxygen ions, showing a dependence on the mass. Altogether, we can conclude that $\langle\Delta E\rangle_{10\%}$ linearly depends on the ion charge. On the other hand, looking at the panel (d), we can notice that the energy gain has a weaker dependence on the ion mass. In particular, $\langle\Delta E\rangle_{10\%}$ is approximately proportional to $ m^{1/5} $ with a $ \chi^{2} < 0.018$. The best fit analysis (dashed lines), indicates that this dependence is weaker than the square root of the mass $ \sqrt{m} $, at variance with what was found in \citet{Greco15}. This is also different from the energy gain expected for non adiabatic orbits in the dawn-dusk electric field $E_{0y}$, for which $\Delta E$ is mass proportional, see \citet{Zelenyi07,Kronberg14}.

\begin{figure}
 	\centering
 	
 	\includegraphics[width=30pc]{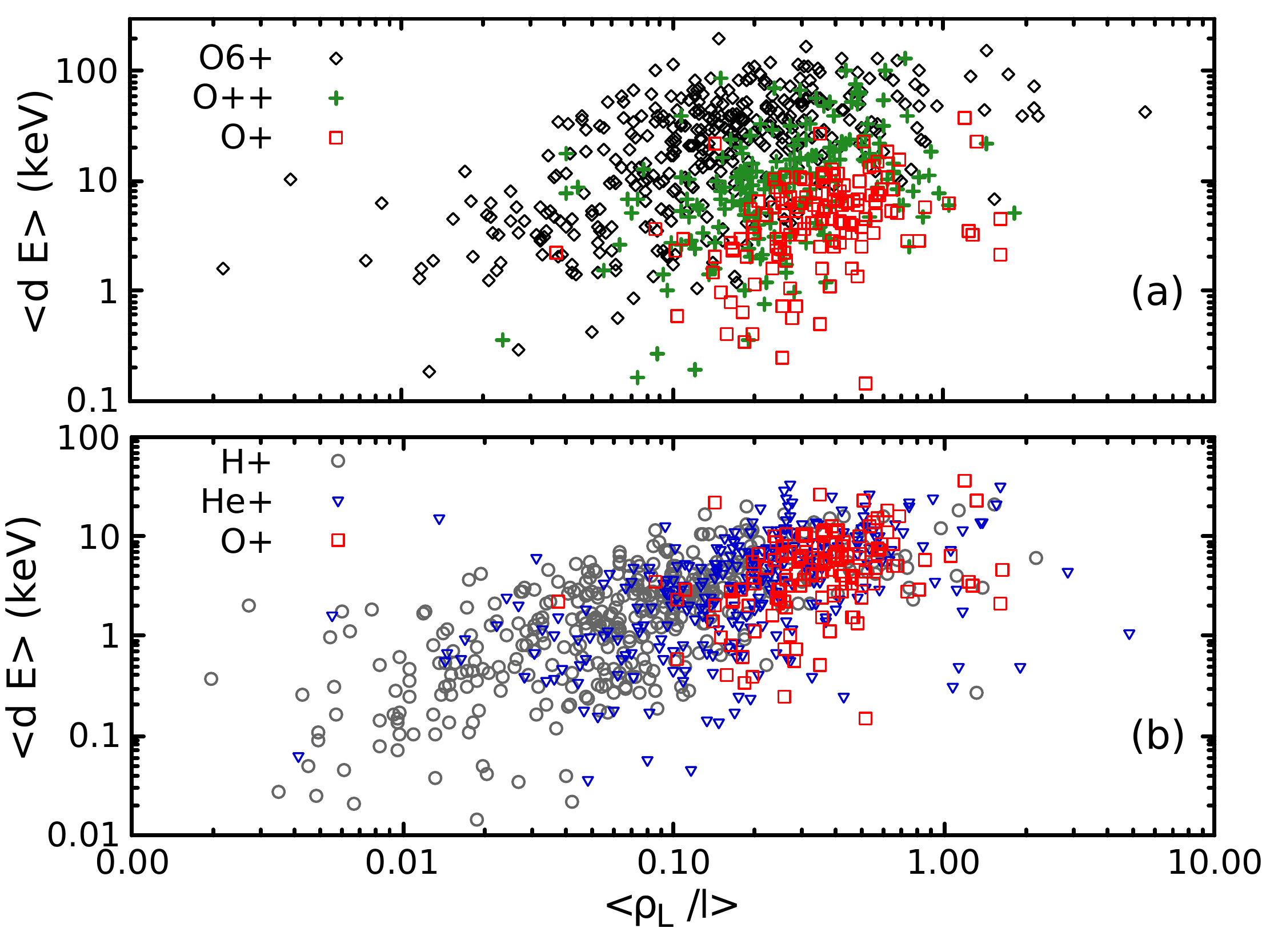}
 	
 	\caption{Scatter-plot of the average energy gain $ \langle dE\rangle $ as function of Larmor radius $\varrho_L$ normalized to $\ell$ for O$^{+}$ (red triangles), O$^{++}$ (green crosses) and O$^{6+}$ (black diamonds) in panel (a), and  H$^+$ (gray circles), He$^+$ (blue triangles) and O$^{+}$ in panel (b).}
 	\label{fig4}
 \end{figure}

We may consider that the energy gain in a single interaction between ions and perturbations could be estimated as $dE = q \delta B V \Delta s$, where $\delta B$ is the amplitude of the magnetic fluctuations, $V$ is the oscillation speed of the perturbations, and $\Delta s$ is the particle displacement within the perturbation fields. If $\Delta s$ would be just proportional to the Larmor radius $\rho_L = m v_{\perp} / q B  $ , a strong mass dependence and no charge dependence should result. If $\rho_L>\lambda$, $\Delta s\propto \lambda$ and the energy gain is only charge proportional; if $\rho_L<\lambda$, $\Delta s\propto \sqrt{\lambda \rho_L}$ \citep{Dob68}, and the energy gain depends on the square root of both charge and mass. Since the model is time dependent, an intermediate scenario occurs. We leave this point to a forthcoming study.
To gain further insights, we compute the local energy gain from the particle integration, and we made a scatter-plot of the average local energy gain $ \langle dE\rangle $ as a function of the local Larmor radius $\rho_L$ normalized to the fluctuations size $\ell$, for $N_p = 500$ particles and for the case with $\ell=8000$ km and $E_{inj} = 1$ keV. The average is made along the particle trajectories, computing $dE$ for each time interval equal to the half gyro-period of the species, but considering only the cases when the particles are in the CS, i.e., $|z| < \lambda$, where most acceleration is expected to occur. The results are shown in Figure \ref{fig4}, where the gray circles denote the H$^+$ ions, blue triangles the He$^+$, red squares the O$^+$, green crosses O$^{++}$ and black diamonds the O$^{6+}$ ions. In all cases there is a growth of the energy gain with the Larmor radius, indicating that ions are more easily accelerated by the fluctuating fields when they are less magnetized, as expected. However, the energy gain proportionally increases with the ion charge, as shown in panel (a). On average, O$^{++}$ and O$^{6+}$ ions have smaller Larmor radii than O$^+$, but the energy gains appear to be larger, confirming the prevailing influence of the ion charge. From panel (b) we can observe that oxygen ions on average have larger Larmor radii, but, compared to the other ion species with the same charge, the energy gain does not increase in proportion (in agreement with the results in Figure \ref{fig3}). This could be due to the fact that because of their large Larmor radius, the O$^+$ ions can exit the CS during their half gyro-period, so that they are not continuously accelerated.

\section{Summary and Discussion}

In this paper we have studied, by means of test particle simulations, the acceleration of different ion species in the Earth's magnetotail in the presence of random electromagnetic perturbations. The equilibrium magnetic and electric field configuration is obtained from the generalized Harris solutions of \citet{Catapano15}, and the 3D electromagnetic perturbations are implemented as in \citet{Perri11}. We injected $ 10^4 $ particles for each run, varying parameters like the ion mass, the ion charge, the initial energy and the perturbation size, and the main results are the following:

1. All ion species are energized up to 100 keV or more. The PDF(E) of all the ions except O$^+$ exhibit non thermal high energy tails. The fraction of time that each species spends in the current sheet ($|z|<\lambda$) was found to decrease with the mass-to-charge ratio $\mu$. Therefore, O$^+$ spends a shorter time in the current sheet and so the efficiency of the Fermi-like acceleration is decreased. This can explain the fact that the O$^+$ energy distribution is almost Gaussian, with no sign of a supra-thermal tail.

2. Considering O$^{n+}$ ions we found that the energy gain grows with the ion charge $n$ in a linear way. Linear fit shows that for the ions with the top 10\% energies, the energy grows as  $\sim 92$ keV per unit charge. In particular, O$^{5+}$ and O$^{6+}$ can reach energies of the order of 500 keV.

3. Assuming virtual particles with charge equal to $q_p$ and the mass equal to 1, 2, 4, 8, and 16 proton masses, we find that the energy gain increases modestly with the mass; a power law fit for the ions with the top 10\% energies yields $\Delta E \simeq (m_i/m_p)^{1/5}$, that is a scaling much slower than other numerical and analytical estimates.

4. We computed the local energy gained along each particle trajectory in the current sheet, and we found that the energy gain increases when the gyroradius is larger. However, the local energy gain does not strongly increase with the mass, that is, O$^+$ ions are accelerated more slowly than other ion species. 

Our results can be useful for comparison with the MMS ion measurements in the magnetotail, specially for those instruments, like FEEPS and EIS, which allow to discriminate the ion mass. We further consider that, according to Figure 3 of \citet{Allen16a}, the abundance of O$^{5+}$ and O$^{6+}$ in the magnetotail from 10 to 20 $R_E$ can be about $1/3$ of the O$^{+}$ abundance. Since our simulations show that these ions can reach energies of 500 keV and more, here we tentatively propose that the observations of strong oxygen energization by spacecraft which do not discriminate the charge state of heavy ions are influenced by the presence of very energetic O$^{5+}$ and O$^{6+}$ ions.

We can compare our simulation results with those of \citet{Greco15}, where the heavy ion acceleration was studied in 2D in the presence of a single dipolarization front. They found that the energy increase is proportional to the square root of the ion mass, even for sodium and sulfur ions; however, \citet{Greco15} showed that this result requires a dipolarization front size larger than the ion gyroradius. Therefore, we can envisage two alternative scenarios for interpreting the energetic ion observations in the magnetotail: when actual O$^+$ is observed to be more energetic than protons and helium, the acceleration mechanism could be related to a single large scale dipolarization front, originating from strong magnetic reconnection typical of the near Earth tail during an intense substorm. When H$^+$ and O$^+$ have more or less the same maximum energy and the proton distribution has a power law tail while oxygen distribution is close to Maxwellian, the acceleration mechanism may be related to transient electromagnetic fluctuations linked to local magnetic reconnection or turbulence, which is typical of the middle tail \citep{Borovsky03,Zimbardo10}, and to small-scale dipolarization fronts that are typical when small-scale transient reconnection regions occur in many locations simultaneously \citep{Lu15}.

 Several observations show that there is no mass dependence of the ion acceleration \citep{Ohtani15}.
In particular, \citet{Ohtani15} found no preferential acceleration of heavy ions at local dipolarization fronts between 10 and 31 $R_E$. \citet{Ohtani15} argue that this is because the background magnetic field may be so weak that not only the O$^+$ ions but also the H$^+$ ions are unmagnetized, so that those ion species may be energized without any particular preference. However, our simulation, in which typical parameters for the magnetotail are used, shows that most of the time the ion gyroradius within the current sheet is smaller than the size of the magnetic perturbation (see Figure \ref{fig4}), so that particles are not completely unmagnetized. Rather, we consider that the reported energization of the different ion species is related to the short time spent by O$^+$ ions inside the current sheet, where the  stochastic acceleration mechanism is effective. 
In a future work we plan to better compare the results from our model with spacecraft observations.

\acknowledgments
Authors are grateful to anonymous referees for insightful suggestions that helped us to improve the manuscript. G. Z. and S. P. acknowledge support from the Agenzia Spaziale Italiana under the contract ASI-INAF 2015-039-R.O " Missione M4 di ESA: Partecipazione Italiana alla fase di assessment della missione THOR".
The authors are grateful to D. Mitchell for fruitful discussions.
 Data supporting all the figures reported in this work come from test particle simulations performed on computers at the Physics Department, University of Calabria. Data can be accessed by writing to the following address: filomena.catapano@unical.it.

\listofchanges

\end{document}